\documentclass[12pt]{iopart}

\usepackage{graphicx}
\usepackage{bm}
\usepackage{amssymb}
\usepackage{color}
\usepackage{eufrak}
\newcommand{\nix}[1]{}
\begin{document}

\title[Optical orientation and spin-dependent recombination]{Optical orientation and spin-dependent
recombination in GaAsN alloys under continuous-wave pumping}

\author{E~L~Ivchenko$^1$, V~K~Kalevich$^1$, A~Yu~Shiryaev$^1$, M~M~Afanasiev$^1$ and
Y~Masumoto$^2$}

\address{$^1$ A.F.~Ioffe Physico-Technical Institute, St.
Petersburg 194021, Russia}
\address{$^2$ Institute of Physics, University of Tsukuba, Tsukuba
305-8571, Japan} \ead{kalevich@solid.ioffe.ru}

\begin{abstract}
We present a systematic theoretical study of spin-dependent
recombination and its effect on optical orientation of
photoelectron spins in semiconductors with deep paramagnetic
centers. For this aim we generalize the Shockley--Read theory of
recombination of electrons and holes through the deep centers with
allowance for optically-induced spin polarization of free and
bound electrons.  Starting from consideration of defects with
three charge states we turn to the two-charge-state model
possessing nine parameters and show that it is compatible with
available experimental data on undoped GaAsN alloys. In the weak-
and strong-pumping limits, we derive simple analytic equations
which are useful in prediction and interpretation of experimental
results. Experimental and theoretical dependencies of the
spin-dependent-recombination ratio and degree of photoluminescence
circular polarization on the pumping intensity and the transverse
magnetic field are compared and discussed.

\end{abstract}

\pacs{72.20.Jv, 72.20.Bh, 72.25.Fe, 78.55.-m}
\submitto{J. Phys: Condens. Matter}
\maketitle

\section{Introduction} \label{sec:1}
In recent years proposals for quantum computation and spintronics
have initiated search for semiconductor nonmagnetic bulk materials
and heterostructures that might be used to realize high values of
electron spin polarization, on the one hand, and its long storage
time, on the other. At low temperatures, this task is successfully
realized by exploiting low spin-relaxation rates of charge
carriers localized in semiconductor quantum wells or confined in
self-organized quantum
dots~\cite{Awschalom,Takagahara,Spin,Kavokin}. With increasing
temperature up to room temperature the free carriers get
delocalized and, in zinc-blende-lattice bulk semiconductors, their
spin relaxation time drastically decreases to values of the order
of 100\,ps or even less~\cite{Spin,OO,Kim2002,Murdin2005} which
limits the electron spin polarization under continuous-wave (cw)
photoexcitation to few percent~\cite{Spin,OO}. The situation can
be partially improved in strongly $n$-doped samples where the
intense electron-electron collisions essentially suppress the
D'yakonov-Perel' mechanism of spin relaxation, as shown
theoretically~\cite{Glazov2001} and
experimentally~\cite{Harley2002}, see also a comprehensive review
article by Wu \etal~\cite{review_Wu}

Recent experimental studies of electron spin dynamics in GaAsN
alloys have revealed strong contradiction with the existing
expectations: extremely high spin polarization, up to 90\%, of the
free electrons and its preservation during $\sim$1\,ns have been
found at room temperature in nonmagnetic undoped GaAsN bulk films
and GaInAsN quantum
wells~\cite{Okayama2003,Egorov2005,JETP1,Lombez2005,
ICPS2006,JETP2,pss2007,obzor2008,QW2010}. In these experiments
electron spin polarization was created by optical pumping and
detected by polarized photoluminescence (PL). It has been
established that the anomalously high polarization and spin memory
arise due to spin-dependent recombination of free electrons with
deep paramagnetic centers present in the nitrogen alloys and
subsequent dynamic spin polarization of the
centers~\cite{JETP1,Lombez2005,JETP2,obzor2008}. The latter acts
as a spin filter preventing the spin majority photoelectrons to
recombine and promoting the spin minority photoelectrons to
disappear from the conduction band. As a result the total density
of photoelectrons increases superlinearly with the increasing
pumping power which leads to an enhancement of the edge PL
intensity (up to a factor of 8)~\cite{JETP1,Int1,apl2009} as well
as of room-temperature photoconductivity~\cite{current,Int2}.
Recent measurements of optically detected magnetic resonance
(ODMR)~\cite{apl2009,naturemat} and optical orientation of
electron spins~\cite{Lombez2005,obzor2008} have provided a proof
that the paramagnetic centers in question are $\rm{Ga}_i$
self-interstitial defects and established the critical dependence
of the defect concentration on the N composition, Ga(In)AsN alloy
growth conditions and post-growth treatment, e.g., thermal
annealing.

The spin-dependent recombination (SDR) gives rise to formation of
a coupled spin system of free and bound electrons where the
free-electron polarization is intricately related to the spin
state of paramagnetic centers. For the quantitative description of
this coupled nonlinear system we proposed the SDR
model~\cite{JETP1,JETP2} based on the model introduced by Weisbuch
and Lampel~\cite{Lampel} and used later by Paget~\cite{Paget} in a
qualitative analysis of low-temperature optical orientation of
electron spins in GaAlAs and GaAs, respectively. Our model
contains nine parameters. At present most of them have been found
for GaAsN experimentally. Particularly, the measurements of cw and
time-resolved polarized luminescence under circularly-polarized
photoexcitation in the presence of a transverse magnetic field
have allowed to estimate the spin-relaxation times of free and
bound electrons, lifetime of photoholes, absolute value of
free-electron $g$ factor and signs of the free- and bound-electron
$g$-factors. The $g$ factor of bound electrons is determined from
ODMR studies~\cite{naturemat}.

In this paper we present a systematic theoretical study of the SDR
in the three- and two-charge-state models with a focus
concentrated on the conduction photoelectron spin polarization and
circular polarization of edge PL induced by circularly-polarized
radiation. It is shown that the two-charge-state theory is
compatible with available experimental data on GaAsN alloys. The
influence and specific role of each model parameters is analyzed
and a complete set of them is determined for a particular GaAsN
alloy sample. The experimental curves are compared not only with
the results of exact model calculation but also with approximate
analytical equations derived in the limit of weak or strong
pumping powers and small or high magnetic fields.

The paper is organized as follows. In section\;\ref{sec:2} we
extend the theory of Shockley-Read recombination to allow for the
spin polarization and spin-dependent recombination of
photocarriers. We start from a more general three-charge-state
model of the SDR and then reduce the consideration to the
two-charge-state model for the particular case of cw
photoexcitation. The exact equations are derived for the PL
intensity and circular polarization, and analytical equations are
derived in the simple limiting cases. In section\;\ref{sec:3}, the
values of nine parameters of the two-charge-state model are
estimated by using available experimental data and reasonable
theoretical considerations. Experimental and theoretical
dependencies of the SDR ratio and degree of PL circular
polarization on the pumping intensity and magnetic field are
presented, compared and discussed in section\;\ref{sec:4}.
Section\;\ref{sec:5} contains the concluding remarks.
\section{Theory} \label{sec:2}
\subsection{Three-charge-state model}

\begin{figure}
  \centering
    \includegraphics[width=0.8\linewidth]{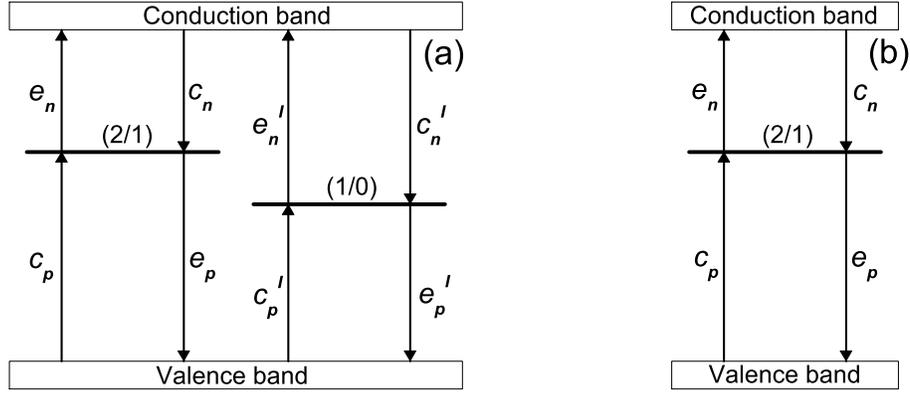}
    \caption{Band diagrams of the three-charge-state
(a) and two-charge-state (b) models. Upper arrows indicate the
excitation (up arrows) and recombination (down arrows) of
conduction electrons described by the coefficients $e_n, c_n$ for
the $(2/1)$ transition between the two- and one-electron defect
states and $e'_n, c'_n$ for the $(1/0)$ transition. Lower arrows
illustrate the hole excitation ($e_p, e'_p$) and recombination
($c_p, c'_p$). }
      \label{Diagr}
      \end{figure}

In solids, impurity atoms can form several energy levels in the
band gap. To illustrate we refer to four-level Au impurity in
germanium~\cite{Au_in_Ge} and three-charged states of interstitial
boron in silicon~\cite{B_in_Si} or Fe, Rh in photorefractive
crystals~\cite{photorefractive,photrefractive2}. The multicharged
states of Ga interstitials in GaAs are analyzed in
reference~\cite{Ginterstitial}, see also references therein. The
statistics of defects with several trapping levels in
semiconductors was analyzed decades ago~\cite{shockley2}. Until
now the rate equations describing light-induced changes in
electron and hole densities and occupation of three-charged
impurity states has been treated for spin-unpolarized charge
carriers~\cite{photorefractive,photrefractive2}. In this section
we generalize the consideration to the spin polarized
photoelectrons and their spin-dependent recombination via deep
paramagnetic centers.

The three-charge-state model is summarized in figure~1(a). We
assume that each deep center can be in one of three states $D_i$
($i=0, 1, 2$) representing the center containing, respectively, no
electrons, one bound electron with uncompensated spin $\pm 1/2$
and two bound electrons in the singlet state with the zero total
spin. The corresponding densities are denoted as $N_0, N_1, N_2$
and satisfy the condition
\begin{equation} \label{deepcenter}
N_0 + N_1 + N_2 = N_c\:,
\end{equation}
where $N_c$ is the total density of the deep-level centers.

Under normal incidence of the circularly polarized light in the
absence of a magnetic field, the electronic spins are polarized
along the excitation direction (the axis $z$).  Let us introduce
the notations $n_{\pm}, N_{\pm}$, respectively, for the
concentration of the free electrons and paramagnetic centers with
the electron-spin component $\pm 1/2$ of the electron spin. Note that $N_+ + N_- = N_1$.
Hereafter, we assume the spin relaxation of photoholes to be very
fast, neglect their spin polarization and use the notation $p$ for
the hole density. Taking into account interband optical generation
of free carriers, their recombination and thermal ionization from
impurity levels, the rate equations for the electron densities
$n_{\pm}$, the hole density $p$ and the densities $N_{\pm}, N_0$
at zero magnetic field are
\begin{eqnarray} \label{maineqs}
\frac{dn_{\pm}}{dt} + \frac{n_{\pm}
- n_{\mp}}{2 \tau_s} + \gamma n_{\pm} p= - R_{n,\pm} + G_{n,\pm} - R'_{n,\pm} + G'_{n,\pm}  + G^{\rm opt}_{\pm}\:,
\\
\frac{dN_{\pm}}{dt} + \frac{N_{\pm} -
N_{\mp}}{2 \tau_{sc}} = \nonumber\\ \mbox{} \hspace{1 cm} - R_{n,\mp} + G_{n,\pm} + R'_{n, \pm} - G'_{n,\pm} + R_{p,\pm} - G_{p, \pm}
- R'_{p,\pm} + G'_{p, \pm} \:,
\nonumber\\
\frac{dp}{dt} + \gamma n_{\pm} p = - R_p - R'_p + G_p + G'_p + G^{\rm opt}\:,\nonumber\\ \frac{d N_0}{dt} =
- R'_n + G'_n + R'_p - G'_p \:. \nonumber
\end{eqnarray}
Here $G^{\rm opt}_{\pm}$ are the photogeneration rates of free
electrons with the spin $\pm 1/2$, $G^{\rm opt}$ is the total
generation rate $G^{\rm opt}_+ + G^{\rm opt}_-$ proportional to the incident light intensity $J$, $\gamma$ is the
coefficient for band-to-band recombination, the times $\tau_s$ and
$\tau_{sc}$ describe the spin relaxation of free and single bound
electrons (apparently, at room temperature $\tau_s \ll
\tau_{sc})$. The rates for electron or hole recombination and
thermal excitation are given by
\begin{eqnarray} \label{maineqs2}
&&R_{n,\pm} = 2 c_n N_{\mp} n_{\pm}\:,\: R'_{n,\pm} = c'_n N_0 n_{\pm}\:,\nonumber\\ &&
R'_{n} = R'_{n,+} + R'_{n, -} = c'_n N_0 (n_+ + n_-)\:, \nonumber \\ &&R_{p} = 2 R_{p, \pm} = c_p N_2 p\:,\:
R'_{p, \pm} = c'_p N_{\pm} p\:,\:\:R'_p = c'_p (N_+ + N_-) p\:, \nonumber\\
&&G_{n,\pm} = \frac{e_n}{2} N_2\:,\: G'_{n,\pm} = e'_n N_{\pm}\:,\:G'_{n} = G'_{n+} + G'_{n-}\:, \nonumber\\
&& G_{p, \pm} = e_p N_{\pm}\:,\:G_p = e_p (N_+ + N_-)\:,\: G'_p = 2 G'_{p, \pm} = e'_p
N_0\:, \nonumber
\end{eqnarray}
the coefficients $c_n, e_n, c_p, e_p$ describe the $(2/1)$
processes $D_1 + e^- \leftrightarrow D_2$ and $D_2 + h^+
\leftrightarrow D_1$ and have the same meaning as the
corresponding coefficients in the original article by Shockley and
Read~\cite{shockley} (see also the book~\cite{AbPeYa}), the
similar coefficients $c'_n, e'_n, c'_p, e'_p$ refer to the $(1/0)$
processes $D_0 + e^- \leftrightarrow D_1$ and $D_1 + h^+
\leftrightarrow D_0$, where $e^-$ and $h^+$ symbolize an electron
and a hole.

In addition to equation~(\ref{deepcenter}) the densities in
consideration satisfy the neutrality equation
\begin{equation} \label{neutr}
p = n + N_2 - N_0  - (N^0_2 - N^0_0)\:,
\end{equation}
where $N^0_j$ are the equilibrium densities given
by~\cite{shockley2}
\[
\frac{N^0_2}{N^0_1} = \frac12 \exp{\left(\frac{\mu - E(2/1)}{k_B
T}\right)}\:,\:~~~ \frac{N^0_1}{N^0_0} = 2 \exp{\left(\frac{\mu -
E(1/0)}{k_B T}\right)}
\]
and $N^0_0 + N^0_1 + N^0_2 = N_c$, where $\mu$ is the electron
equilibrium chemical potential, $E(i+1/i)$ is the change in energy
associated with adding one electron to the center in the process
$D_i + e^- \rightarrow D_{i+1}$ and the factors 2 and 1/2 take
into account the double spin degeneracy of the paramagnetic state.

The set of equations~(\ref{deepcenter})--(\ref{neutr}) contains a
lot of parameters. To simplify the analysis we assume that the
photoexcitation rate $G^{\rm opt}$ is high enough to neglect the
thermal excitation rates $G_{n,\pm}, G'_{n,\pm}, G_p$ and $G'_p$.
Moreover, we neglect the influence of band-to-band recombination
on the kinetics of charge carriers, set $\gamma$ to zero in
equations~(\ref{maineqs}) and use the thus obtained values of
$n_+, n_-$ and $p$ to calculate the photoluminescence (PL)
intensity and polarization. For the sake of distinctness, we also
assume that, in equilibrium, all the centers bind one electron per
center, i.e., are paramagnetic and $N_1^0 = N_c$.

Instead of four variables $n_{\pm}, N_{\pm}$ we introduce the
total free-electron density $n= n_+ + n_-$ and spin $S_z= (n_+ -
n_-)/2$, the density of paramagnetic centers (or centers with
uncompensated spins) $N_1 = N_+ + N_-$ and the bound-electron
total spin $S_{c z}= (N_+ - N_-)/2$. In a transverse magnetic
field the electron spin exhibits the Larmor precession and,
instead of the scalars $S_z$ and $S_{c z}$, one should define the
pseudovectors ${\bm S}$ and ${\bm S}_c$ pointing in the directions
of the average free- and bound-electron spins. Then the set of
kinetic equations (\ref{maineqs}) in the presence of magnetic
field are rewritten as \numparts \label{7II}
\begin{eqnarray}
\frac{dn}{dt} + c_n (N_1 n - 4 {\bm S} {\bm S}_c) +
c'_n N_0 n = G^{\rm opt}\:, \label{7IIa}\\
\frac{dp}{dt} + (c_p N_2 + c'_p N_1)p = G^{\rm opt}\:, \label{7IIb}\\
\frac{d{\bm S}}{dt} + c_n (N_1 {\bm S} - {\bm S}_c n) + {\bm S}
\left(\frac{1}{\tau_s} + c'_n N_0 \right) + {\bm S} \times {\bm
\omega}= \frac{P_i}{2}\ G^{\rm opt} {\bm o}_z\:, \label{7IIc}
\\ \frac{d {\bm S}_c}{dt} + c_n ({\bm S}_c n -  N_1 {\bm S})
+ \frac{{\bm S}_c}{\tau_{sc}} - c'_n N_0 {\bm S} + c'_p p {\bm S}_c + {\bm S}_c \times
{\bm \Omega}= 0\:, \label{7IId}\\
\frac{dN_0}{dt} + c'_n N_0 n + c'_p N_1 p =  0\:. \label{7IIe}
\end{eqnarray}
\endnumparts
Here $P_i = (G^{\rm opt}_+ - G^{\rm opt}_-)/G^{\rm opt}$, ${\bm
o}_z$ is the unit vector directed along the normal $z$ coinciding
with the exciting-light propagation direction, ${\bm \omega}$ and
${\bm \Omega}$ are the Larmor frequencies defined by $\hbar {\bm
\omega} = g \mu_B {\bm B}$, $\hbar {\bm \Omega} = g_c \mu_B {\bm
B}$, $g$ and $g_c$ are the $g$-factors of free and bound
electrons, ${\bm B}$ is the external magnetic field and $\mu_B$ is
the Bohr magneton. In the following we assume ${\bm B}
\parallel y \perp z$. Equations~(\ref{7IIa})--(\ref{7IIe}) together with
equations~(\ref{deepcenter}) and (\ref{neutr}) form a complete set
in the proposed model.
\subsection{Continuous wave photoexcitation}
Under continuous wave photoexcitation the time derivatives vanish
and the equations for ${\bm S}$ and ${\bm S}_c$ are reduced to
\begin{eqnarray} \label{sS}
&& \frac{\bm S}{T} - \frac{{\bm S}_c}{\tau_c}
+ {\bm S} \times {\bm \omega} = \frac{P_i}{2}\ G^{\rm opt} {\bm o}_z\:, \\
&& \frac{{\bm S}_c}{T_c} - \frac{\bm S}{\tau} + {{\bm S}_c} \times
{\bm \Omega}= 0\:,\nonumber
\end{eqnarray}
where the four introduced times are defined by
\begin{eqnarray} \label{Sigma}
&& \frac{1}{T} = \frac{1}{\tau_s} + \frac{1}{\tau} \:,\hspace{1.9 cm}\frac{1}{\tau_c} = c_n n\:, \\
&& \frac{1}{T_c} = \frac{1}{\tau_{sc}} + c_n n + c'_p p
\:,\:\:\:\: \frac{1}{\tau} = c_n N_1 + c'_n N_0\:. \nonumber
\end{eqnarray}
By using equations~(\ref{sS}) we can present the free- and
bound-electron spin components $S_z, S_{cz}$ in the form
\begin{equation} \label{szpsi}
S_z = \frac{P_i G^{\rm opt} T}{2 \Psi}\:,\: \hspace{10 mm}S_{cz} =
\Lambda S_z\:.
\end{equation}
where
\begin{equation} \label{psi}
\Psi = 1 + \omega^2 \tilde{T}^2 - \eta \frac{(1 - \omega \Omega
\tilde{T} \tilde{T}_c)^2}{1 + \Omega^2 \tilde{T}_c^2}\:,
\end{equation}
\[
\Lambda = \frac{T_c}{\tau}\ \frac{1 - \omega \Omega \tilde{T}
\tilde{T}_c}{1 + \Omega^2 \tilde{T}_c^2}\:,
\]
and $\eta = (TT_c/\tau \tau_c)$,~ $\tilde{T} = T/\sqrt{1 -
\eta}$,~ $\tilde{T}_c = T_c/\sqrt{1 - \eta}$. The transverse spin
components $S_x, S_{cx}$ (for ${\bm B} \parallel y$) are given by
\begin{equation} \label{sxscx}
S_x = \frac{T}{\tau_c} \frac{\omega \tau_c S_z + \Omega T_c
S_{cz}}{1 - \eta}\:,\hspace{10 mm}S_{cx} = \frac{T_c}{\tau}
\frac{\omega T S_z + \Omega \tau S_{cz}}{1 - \eta}\:.
\end{equation}

The densities $p, N_0, N_2$ can be expressed via $n$ and $N_1$ as
\begin{eqnarray}
&&p = \frac{c'_n n}{2 c'_p N_1} \frac{N_c - N_1 + n}{1 +
(c'_n/2 c'_p)(n/N_1)}\:, \\
&&N_0 = \frac{c'_p N_1}{c'_n n} p = \frac12 \frac{N_c - N_1 + n}{1 + (c'_n/2 c'_p)(n/N_1)} \:, \nonumber\\
&&N_2 = \frac12 \frac{(N_c - N_1) [1 + (c'_n/c'_p)(n/N_1)] - n}{1
+ (c'_n/2 c'_p)(n/N_1)}\:.
\end{eqnarray}

Excluding $p, N_0, N_2$, ${\bm S}$ and ${\bm S}_c$ from the
complete set of equations we obtain, instead of
equations~(\ref{7IIa}) and (\ref{7IIb}), the following two coupled
nonlinear equations for $n$ and $N_1$
\begin{eqnarray} \label{long2}
\frac12 \left\{ c_p \frac{(N_c - N_1) [1 + (c'_n/c'_p)(n/N_1)] -
n}{1 + (c'_n/2c'_p)(n/N_1)} + 2 c'_p N_1 \right\}  \nonumber\\
\mbox{} \hspace{13 mm}\times \frac{N_c - N_1 + n}{(2 c'_p/c'_n)(N_1/n) + 1} = G^{\rm opt}\:, \\
\label{long}
c_n n N_1 ( 1 - \Sigma ) + c'_n N_0(n,N_1) n = G^{\rm opt}\:,\hspace{3 cm} \mbox{}
\nonumber
\end{eqnarray}
where
\begin{eqnarray} \label{sigmapsi}
\Sigma = \frac{4 {\bm S} {\bm S}_c}{n N_1} = \frac{(P_i G^{\rm opt}
T)^2}{n N_1 \Psi^2}\\ \mbox{} \hspace{8 mm} \times  \left\{ \Lambda + \frac{\eta}{(1 - \eta)^2} [\omega^2
\tau_c T + \Lambda \omega \Omega \tau \tau_c (1 + \eta) +
\Lambda^2 \Omega^2 \tau T_c ] \right\} \:. \nonumber
\end{eqnarray}

According to equation~(\ref{szpsi}) the degree of free-electron
spin polarization is given by
\begin{equation} \label{spinpol}
P = \frac{2S_z}{n}= \frac{P_i G^{\rm opt} T}{n \Psi}\:.
\end{equation}
The intensity $I$ and degree  of circular polarization of the
photoluminescence $\rho$ can be found from
\begin{equation} \label{intpol}
I = I^+ + I^- \propto n p\:,\: \hspace{15 mm}\rho = \frac{I^+ -
I^-}{I} = P' P \:,
\end{equation}
where $I^{\pm}$ are the intensities of the $\sigma^{\pm}$
circularly-polarized components of the interband PL, and $P' =
P'_1 P'_2$, with the factor $P'_1$ taking into account the
selection rules for the room-temperature recombination of the
conduction electrons with heavy and light holes, and $P'_2 \leq 1$
being the depolarizing factor arising due to possible multiple
reflections from the sample boundaries~\cite{asnin}.
\subsection{Two-charge-state model}
As follows from
references~\cite{JETP1,ICPS2006,JETP2,pss2007,obzor2008,naturemat,physica}
and demonstrated below the SDR model taking into account only the
deep-center states $D_1$ and $D_2$ satisfactorily describes the
experiments on spin dynamics in GaAsN samples. Therefore, in what
follows we focus on this particular model setting $c'_p = 0$ in
which case $N_0 =0$, $N_c = N_1 + N_2$, $p = n +  N_c - N_1$ and
equations~(\ref{long2}), (\ref{long}) reduce to
\begin{equation} \label{reduced}
c_p (N_c - N_1) (N_c - N_1 + n) = G^{\rm opt},~~~\:c_n n N_1 ( 1 -
\Sigma )  = G^{\rm opt}\:.
\end{equation}

{\it Optical orientation of electron spins at zero magnetic field}.
Let us consider particular limiting cases allowing for simple
analytical expressions. In the absence of zero magnetic field the
Larmor frequencies $\omega, \Omega $ vanish and the expressions
for $S_z$ and $\Sigma$ reduce to
\begin{equation} \label{szsigma}
S_z = \frac{P_i G^{\rm opt} T}{2 (1 - \eta)}\:,\: ~~~~\Sigma =
\frac{(P_i G^{\rm opt} T)^2}{n N_1} \frac{T_c}{\tau} \frac{1}{(1 -
\eta)^2}\:.
\end{equation}
It is convenient to introduce the dimensionless variables
\begin{equation} \label{XYZ}
X = \frac{G^{\rm opt}}{c_p N_c^2}\:,\: ~~~~Y = \frac{N_c -
N_1}{N_c} = \frac{N_2}{N_c}\:,\: ~~~~Z = \frac{n}{N_c}\:.
\end{equation}
In terms of these variables one has
\begin{eqnarray} \label{pIrho}
p = N_c (Y + Z)\:,\:\hspace{3 mm} I \propto Z(Y + Z)\:,\:\hspace{3 mm} \rho = \frac{P'
P_i}{1 - \eta}~\frac{T}{\tau^*_h}~\frac{X}{Z}\:, \\
\Lambda = \frac{T_c}{\tau}\:,\:\hspace{7 mm} \Psi = 1 - \eta\:, \nonumber\\
\frac{1}{\tau} =
\frac{1-Y}{\tau^*}\:,\:\hspace{5 mm} \frac{1}{T} = \frac{1}{\tau_s} +
\frac{1-Y}{\tau^*}\:, \nonumber \\ \frac{1}{\tau_c} =
\frac{Z}{\tau^*}\:,\:\hspace{10 mm} \frac{1}{T_c} = \frac{1}{\tau_{sc}} +
\frac{Z}{\tau^*}\:,\nonumber
\end{eqnarray}
where $\tau^* = (c_n N_c)^{-1}$ and $\tau^*_h = (c_p N_c)^{-1}$.

According to equations~(\ref{long}) and (\ref{reduced}) the
variables $Y$ and $Z$ satisfy the equations
\begin{eqnarray} \label{XYZequation}
Y(Y+Z) = X \:, \\ \frac{1-Y}{a} \left\{ Z - P_i^2 \left(
\frac{\tau_s}{\tau_h^*} \right)^2 X^2 \frac{Z +
\tau^*/\tau_{sc}}{[Z + \tau^*/\tau_{sc} + (1-Y)
\tau_c/\tau_{sc}]^2} \right\} = X\:, \nonumber
\end{eqnarray}
where $a = c_p / c_n$.

In the low-power limit we have $\tau_c \to \infty, ~\eta \to 0$
and
\begin{eqnarray} \label{gp'}
n = \tau^* G^{\rm opt}\:,\hspace{7 mm}\rho = P' P_i \frac{\tau_s}{\tau_s +
\tau^*}\:, \\ p \propto N_c - N_1 \propto \sqrt{G^{\rm
opt}}\:,\hspace{5 mm} I \propto (G^{\rm opt})^{3/2}\:. \nonumber
\end{eqnarray}
The first nonvanishing corrections to $n$ and $\rho$ are taken
into account in equation~(\ref{gp'}) by the replacement $\tau^*
\to \tau^*(1 + \sqrt{\tau^*_hG^{\rm opt}/N_c}\ )$. In particular,
this means that
\begin{equation} \label{lowinta}
\rho = P' P_i \frac{\tau_s}{\tau_s + \tau^*} \left( 1 -
\frac{\tau^*}{\tau_s + \tau^*} \sqrt{\frac{\tau^*_hG^{\rm opt}}{N_c}}
\right)\:.
\end{equation}
However, as soon as a value of $(\tau_s \tau_{sc}/\tau^*
\tau_h^*)\sqrt{X}$ exceeds unity the main correction to $\rho$
originates from the spin-dependent recombination, it is linear in
$G^{\rm opt}$ and given by
\begin{equation} \label{lowintb}
\rho = P' P_i \frac{\tau_s}{\tau_s + \tau^*} \left( 1 +
\frac{\tau_s}{\tau_s + \tau^*}
\frac{\tau_{sc}G^{\rm opt}}{N_c} \right)\:.
\end{equation}

With increasing pump intensity, $Y$ monotonously increases and
saturates up to some limiting value controlled by the ratio
$c_p/c_n$. This limiting value $Y_{\infty} \equiv Y( G^{\rm opt}
\to \infty)$ satisfies a third-order algebraic equation
\begin{equation} \label{yinfty}
(1 - Y_{\infty})(1 - b Y_{\infty}^2) - a Y_{\infty} = 0\:,
\end{equation}
where $b = P_i^2 (\tau_s/\tau^*_h)^2$. Note that according to
equation~(\ref{XYZequation}) one has asymptotically $Z(G^{\rm opt}
\to \infty) \to X/Y_{\infty}$. If $a = c_p / c_n \ll | 1 -
\sqrt{b} |$, then
\begin{equation} \label{yinfty2}
Y_{\infty} \approx \left\{ \begin{array}{c} \hspace{4 mm} 1 - \frac{a}{1 - b} \hspace{15 mm} \mbox{for}
\hspace{2 mm} b < 1\:, \\
\frac{1}{\sqrt{b}}  \left( 1 -
\frac{a}{2(\sqrt{b} - 1)} \right) \hspace{2 mm} \mbox{for} \hspace{2
mm} b > 1 \:.  \end{array} \right.
\end{equation}
For $b \approx 1$ and $a \ll 1$, one has $Y_{\infty} = 1 - \sqrt{a/2}$.

{\it Hanle effect at transverse magnetic field}.
Assuming $T \ll T_c$ and comparable values of $g$ and $g_c$ we can separately analyze two
regions of the transverse magnetic field:
(I) the low-field region where $\omega T \ll 1$ and the product $\Omega T_c$
is arbitrary, and (II) the high-field region where
$\Omega T \gg 1$ and the product $\omega T$ is arbitrary.

In the region I, at low pumping level, equation (\ref{lowintb})
derived at $B=0$ transfers to
\begin{equation} \label{lowintB}
\rho(B) = P' P_i \frac{\tau_s}{\tau_s + \tau^*} \left( 1 +
\frac{\tau_s}{\tau_s + \tau^*} \frac{\tau_{sc} G^{\rm opt} }{N_c} \frac{1}{1 + \Omega^2 T_c^2}  \right)\:.
\end{equation}
One can see that in this particular case $\rho(B)$ is a sum of a constant and a weak narrow Lorentzian.

In the region II, at arbitrary pumping level, one can neglect the
influence of bound electrons upon the formation of the
free-electron density and polarization, set $\Sigma$ to zero and
use the equations
\begin{equation} \label{arbint}
n = \tau G^{\rm opt} \:,\:~~~ \rho(B) = P' P_i
\frac{\tau_s}{\tau_s + \tau} \frac{1}{1 + \omega^2 T^2} =
\frac{\rho_0}{1 + (B/B_{1/2})^2}\:,
\end{equation}
where $\rho_0 = P' P_i \tau_s/(\tau_s + \tau)$ and $B_{1/2} = \hbar/(g \mu_B T)$ is the half-width at
half-maximum of the Lorentzian. At weak incident-light intensities, the lifetime $\tau$ can be replace
by the characteristic time $\tau^*$.

{\it Electron density for unpolarized carriers}. The system is
spin-unpolarized under linearly-polarized
photoexcitation as well as for $\sigma^{\pm}$ pumping in the limit
of very strong transverse magnetic fields. In the both cases, in order to find $n$
and $N_1$ one can formally set $P_i = 0$ and reduce a pair of equations
(\ref{XYZequation}) to
\begin{equation} \label{pizero}
\frac{Y^2(1 - Y)}{1 - (1+a)Y} = X\:.
\end{equation}
To trace the saturation of $Y$ and asymptotic behavior of $Z$ with
the increasing pump intensity, we set $X \to \infty$ in
equation~(\ref{pizero}) and obtain
\begin{equation}
Y_{\infty} \approx \frac{1}{1 + a} \left( 1 - \frac{a}{(1 +
a)^3} \frac{N_c}{\tau^*_h G^{\rm opt}}  \right)\:,
\end{equation}
\[
Z( G^{\rm opt} \to \infty) \approx  \frac{(1 + a) \tau^*_h G^{\rm opt}}{N_c}  -
\frac{1}{(1 + a)^2} \:.
\]

\section{Selection of model parameter values} \label{sec:3}
The two-charge-state model is characterized by two free
(controllable) variables, the total generation rate $G^{\rm opt}$
and the strength of magnetic field $B$, and nine fitting
parameters, namely, Land\'e factors $g$ and $g_c$, the spin
relaxation times $\tau_ s$ and $\tau_{sc}$, the recombination
times $\tau^* = (c_n N_c)^{-1}$ and $\tau^*_h = (c_p N_c)^{-1}$,
the degrees of polarization $P_i, P'$ and the density of deep
centers $N_c$. In the present section we will estimate values of
these parameters by using available experimental data and
reasonable theoretical considerations.
\begin{center}
{\it 1. Lande $g$-factors of free and bound electrons}
\end{center}
The recent studies~\cite{apl2009,naturemat} of ODMR have provided
unambiguous experimental evidence that, in dilute nitrides
GaAs$_{1-x}$N$_x$ and Ga$_{1-y}$In$_y$As$_{1-x}$N$_x$ with a few
percent of nitrogen ($x \leq 3.3$\%), the deep paramagnetic
centers are formed by Ga$_i$ self-interstitials. In the above
notations, the defect state $D_1$ occupied by a single electron
represents the doubly positive charged level Ga$_i^{2+}$ while the
singlet $D_2$ occupied by a pair of electrons is identified as the
singly positive state Ga$_i^{1+}$. The bound-electron $g$ factor
$g_c$ determined from the ODMR measurements has been found to be
very close to 2. The positive sign of $g_c$ has been independently
determined from asymmetry in the depolarization of edge
photoluminescence in a transverse magnetic field (Hanle effect) at
oblique incidence of the exciting radiation and oblique-angle
detection of the luminescence~\cite{JETP1,SST2008}. The results of
references~\cite{JETP1,apl2009,naturemat,SST2008} are in agreement
with the existing expectation that the $g$ factors of electrons
bound to deep levels, as a rule, exhibit only a small
renormalization with respect to the gyromagnetic factor $g_0 =
2.0023$ of a free electron in vacuum.

On the other hand, the $g$ values of free carriers, electrons and
holes, in semiconductors with small and moderate band gaps can
drastically differ from $g_0$. Particularly, for conduction-band
electrons  in bulk GaAs the $g$ factor equals $-0.44$ at helium
temperature. In GaAs$_{1-x}$N$_x$ with a few percent of N the
value of electron $g$ factor shifts upwards. The experimental
study~\cite{Reilly2,Reilly3} of the Zeeman splitting of $\sigma^+$
and $\sigma^-$ circularly polarized components of the
photoluminescence spectra of GaAs$_{1-x}$N$_x$ ($x \leq 0.6\%$)
have shown that the conduction-electron $g$ factor exhibits a sign
reversal from negative to positive at $x \approx 0.04\%$ and
increases abruptly up to 0.7 in a very narrow compositional window
between $x = 0.04\%$ and $x=0.1\%$. For $0.1\%<x \leq 0.6\%$, the
dependence $g(x)$ has a not well-defined behaviour and fluctuates
around 0.7 for the highest $x$ values. The striking difference
from the value $-0.44$ is attributed to the resonant defect level
in the conduction band introduced by nitrogen and the repulsion
between the GaAs host matrix conduction band edge and a higher
lying band of localized N resonant states (the so-called
band-anticrossing model introduced by Shan, Walukiewicz
\etal~\cite{BAM}, see also~\cite{SkierbPRB2005}). The agreement
between experimental data and theoretical calculation is
remarkably improved in the generalized band-anticrossing model
which takes into account the interaction of the conduction-band
minimum with localized resonant states formed by nitrogen-atom
pairs and clusters~\cite{Reilly2,Reilly3}.

The positive sign of \emph{g} in GaAs$_{1-x}$N$_x$ film with $x =
2.1\%$ was found at room temperature from the Hanle effect
asymmetry under oblique excitation and detection~\cite{SST2008}
while its absolute value $|g| \approx0.9$ follows from the
electron spin quantum beats in photoluminescence recorded in a
transverse magnetic field after the pulse excitation~\cite{JETP2}.
The data are in agreement with the value $g\approx +1$ estimated
in reference~\cite{SST2008} from the relationship between \emph{g}
and electron effective mass $m^*\approx 0.14m_0$ measured for
$x=2\%$ in reference~\cite{ital}. The positive free-electron
\emph{g}-factor $g=0.97$ was measured in GaAs$_{1-x}$N$_x$/GaAs
quantum well with $x=1.5\%$ by time resolved Kerr rotation at
220\,K~\cite{China2010}. On the other hand, in
reference~\cite{Zhao2009} only a slow decrease of $|g|$ from 0.28
to 0.22 for $x$ increasing from $0.09\%$ to $0.9\%$ has recently
been reported from time-resolved Kerr rotation measurements in
GaAs$_{1-x}$N$_x$ films at room temperature. The difference with
the above values may be related to variations in nitrogen-atom
spatial distribution inside dilute nitride samples studied by the
two groups.

Below, we will use $g = +1$, $g_c =+2$.
\begin{center}
{\it 2. Characteristic free electron and hole lifetimes $\tau^*$ and $\tau^*_h$}
\end{center}
According to equation~(\ref{7IIb}), in the two-charge-state model
($c'_p=0$) the hole recombination rate equals $c_p N_2p$. In the
limit of high pump intensities, $G^{\rm opt} \to \infty$, the hole
lifetime $\tau_h \equiv (c_p N_2)^{-1}$ tends to a limit of
$\tau^*_h = (c_p N_c)^{-1}$. The regime of high pumping was
realized by Kalevich \etal \cite{JETP2} under pulsed optical
excitation of GaAs$_{0.979}$N$_{0.021}$ alloy at room temperature.
The ratio of the photoelectron density $n$ and $N_c$ exceeded 10
and the photoluminescence decay time at the initial stage, (15.0
$\pm$ 0.5)\,ps, could be identified with $\tau^*_h/2$. Thus, in
this particular sample, the time $\tau^*_h$ can be taken equal to
30\,ps or a little bit shorter because, as the analysis shows, in
the experimental conditions a maximum value of $N_2$ was slightly
smaller than $N_c$.

In the low intensity limit, where $\tau \to \tau^*$, one can use
equations~(\ref{gp'}). Moreover, it follows from the
experiment~\cite{JETP2} that the three characteristic times
$\tau^*, \tau^*_h$ and $\tau_s$ satisfy the hierarchy: $\tau^* \ll
\tau^*_h \ll \tau_s$. Therefore, at low pumping levels the
lifetime, $T$, of free-electron spin polarization given by the
first equation\,(\ref{Sigma}) is close to $\tau^*$. This allowed
us to determine $\tau^*$ from measurements of the Hanle effect at
low incident light intensity $J$. The interpolation of the
experimental dependence $B_{1/2}(J) \approx \hbar/(g \mu_B \tau)$
of the free-electron Hanle curve half-width to the limit $J \to 0$
yields~\cite{physica}, for $g = 1$, the conduction electron
lifetime $\tau^* = \tau(J \to 0) = 1.9$\,ps.

For the model calculations we will use $\tau^* = 1.9$ ps and $\tau^*_h = 27$ ps.
\begin{center}
{\it 3. Spin relaxation times of free and bound electrons}
\end{center}
The spin relaxation time $\tau_s$ of free photoelectrons was
found~\cite{ICPS2006,JETP2} from the decay curves of
$$
S_z(t) \propto \frac{I^+(t) - I^-(t)}{p(t)} \propto \frac{I^+(t) - I^-(t)}{\sqrt{I(t)}}
$$
and, independently, from the PL intensity decay, $I(t)$, at the
second stage of the spin-dependent electron dynamics, the partial
intensities $I^{\pm}$ being introduced in equation~(\ref{intpol}).
The obtained two values $(150\pm15)$\,ps and $(144\pm4)$\,ps,
respectively, coincide within the uncertainty limits, confirming
the interpretation of the experimental data. The spin-Kerr
rotation measurements~\cite{Zhao2009} performed on three GaAsN
samples at room temperature give for the free-electron spin
lifetime $T$ the values of 60, 120 and 125\,ps which can serve as
a lower bound for the spin relaxation time $\tau_s$ in these
samples in agreement with the above two values of $\tau_s$.

The spin-relaxation time of bound electrons, $\tau_{sc}$, can be
deduced from measurements of the Hanle depolarization
curve~\cite{JETP1,physica}, or, to be more precise, the narrow
part of this curve related to the depolarization of bound
electrons. Indeed, according to equation~(\ref{lowintB}) the
half-width of the narrow Lorentzian is given by $B_{1/2}^c=\hbar /
g_c \mu_B T_{c}$ which reduces, at $J\to 0$, to $\hbar / g_c \mu_B
\tau_{sc}$. By using the values $B_{1/2}^c \approx 85$\,G and
$g_c=+2$, we obtained~\cite{physica} $\tau_{sc}\approx700$\,ps.

Later on we will use $\tau_s = 140$\,ps, ~$\tau_{sc} =700$\,ps.
\begin{center}
{\it 4. Initial photoelectron spin polarization $P_i$ and the factor $P'$ relating $\rho$ with $P$}
\end{center}
The determination of the initial spin polarization $P_i$ from
experiments on polarized photoluminescence can be complicated by
the presence of a small additional contribution of unclear origin
to the secondary emission. For the circularly-polarized incident
light, this contribution is also polarized and its circular
polarization is insensitive to the transverse magnetic field up to
very high fields. Taking into account the additional contribution
the measured degree of circular polarization is presented as
$\rho_{\rm exp} = (1 - j) \rho + j \rho_{\rm add}$, where $\rho$
is given by equation~(\ref{intpol}), $\rho_{\rm add}$ is the
polarization of the additional radiation and  $j$ is its fraction
in the total intensity of the secondary radiation. Assuming $j$ to
be small, we can approximate $\rho_{\rm exp}$ to
\begin{equation} \label{rhoexp}
\rho_{\rm exp} = \rho + \rho_{\rm res}\:,
\end{equation}
where $\rho_{\rm res} = j \rho_{\rm add}$ is the effective
``residual'' polarization. By using equation~(\ref{rhoexp}) one
can find $P_i$ from the following equation
\begin{equation} \label{Pi2}
P_i \approx  \frac{\rho_{\rm exp}(B=5\,\textrm{kG}) - \rho_{\rm res}}{\rho_{\rm exp}(B=0,{\rm high}~J)-\rho_{\rm res}}
\, P(B=0,{\rm high}~J)\:.
\end{equation}
Here $\rho_{\rm exp}(B=5\,\textrm{kG})$ is the degree of circular
polarization measured at the moderate value of the transverse
magnetic field, $B$=5\,\textrm{kG}, high enough in order to
neglect the spin polarization of bound electrons and apply
equation~(\ref{arbint}) and, on the other hand, low enough to
satisfy the condition $\omega T \ll 1$; $\rho_{\rm exp}(B=0,{\rm
high}~J)$ is the polarization degree measured at zero magnetic
field and at a high photoexcitation level where the spin-filter
effect results in almost 100\% spin polarization of the free
photoelectrons, $P(B=0,{\rm high}~J) \approx 1$. It follows then
from equation~(\ref{intpol}) that
\begin{equation} \label{Pprime}
P' = \frac{\rho_{\rm exp}(B=0, {\rm high}~J) - \rho_{\rm res}}{P(B=0,{\rm high}~J)}\:.
\end{equation}
Furthermore, according to equation~(\ref{arbint}), at the moderate
magnetic field one has $\rho_{\rm exp}(B=5\,\textrm{kG}) = P' P_i
\tau_s/(\tau + \tau_s) + \rho_{\rm res}$. In the studied GaAsN
alloys, at the moderate intensities, the free-electron lifetime is
much shorter than their spin relaxation time and $\rho_{\rm
exp}(B=5\,\textrm{kG})$ reduces to $P' P_i + \rho_{\rm res}$ which
leads to $P_i = [\rho_{\rm exp}(B=5\,\textrm{kG}) - \rho_{\rm
res}]/P'$ and finally to equation~(\ref{Pi2}).

\begin{center}
{\it 5. Density of deep centers $N_c$}
\end{center}
The density of deep centers responsible for the spin-dependent
recombination can be found by fitting the measured dependencies of
the PL intensity and polarization on the incident light intensity
with computation of the kinetic
equations~(\ref{7IIa})--(\ref{7IIe}). The inhomogeneous terms in
these equations are proportional to the generation rate $G^{\rm
opt}$ which can be expressed in terms of the excitation photon
energy $E_{\rm{ph}}$, the incident light intensity $J$, the light
absorption coefficient $\alpha$ and the beam spot area on the
sample surface, $S^*$. Taking into account that the thickness of
the GaAsN layer $L$ is smaller than $\alpha^{-1}$  and assuming
that each photon creates one carrier pair we have
\begin{equation} \label{Dopt}
G^{\rm{opt}} = \frac{\alpha J}{E_{\rm{ph}}S^*}\:.
\end{equation}

Let us estimate a typical value of $G^{\rm {opt}}$ in our
experiments performed on GaAs$_{0.979}$N$_{0.021}$ at room
temperature and the energy $E_{\rm ph}=1.312$\,eV of the exciting
photons. For this nitrogen content the room-temperature band gap
of the alloy equals $E_g \approx1.106$\,eV~\cite{Egorov2005}, and,
therefore, the excess photon energy $E_{\rm ph} - E_g = 206$\,meV.
We take $\alpha = 2 \times 10^4 \, \textrm{cm}^{-1}$ for this
difference $E_{\rm ph}-E_g$ in accordance to values $2.4\times10^4
\, \textrm{cm}^{-1}$  and  $1.6 \times 10^4 \, \textrm{cm}^{-1}$
measured in the alloys with $x=5\%$~\cite{Tu1977} and
$x=2.3\%$~\cite{Tisch2002} in the wide range of photon energies.
For the light spot diameter 0.2\,mm and the intensity $J=1$\,mW
the generation rate $G^{\rm{opt}}$ is $3 \times
10^{23}$\,$\rm{cm}^{-3} \rm{s}^{-1}$.

\begin{figure}
  \centering
    \includegraphics[width=0.4\linewidth]{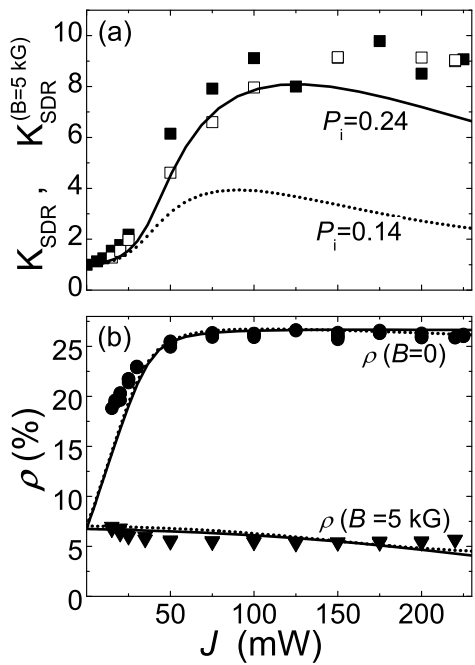}
    \caption{ Spin-dependent recombination ratios (a) and the photoluminescence circular
    polarization (b) in GaAs$_{0.979}$N$_{0.021}$ vs. the excitation power for the incident-photon
    energy $\hbar\omega_{\rm exc}=1.312$\,eV, the detection energy $\hbar\omega_{\rm det} = 1.17\,$eV
    and room temperature. Full and open squares stand for the measured values
    of $K_{\rm SDR} = I({\rm circ})/I({\rm lin})$ and
    $K_{\rm SDR}^{(B= 5\,{\rm kG})} = I({\rm circ}, B=0)/I({\rm circ}, B= 5\,{\rm kG})$, respectively.
Circles and triangles present values of the degree of circular
polarization measured in the absence of magnetic field and at $B =
5$\,kG. Solid and dotted lines are the result of the model
calculation (see text for details).}
    \label{SDRratio}
\end{figure}

A lower estimate for the critical value of the generation rate
$G^{\rm{opt}}_{\rm cr,\,circ}$ at which the spin filter effect
becomes remarkable can be obtained by setting
$\Sigma(G^{\rm{opt}})$ to unity. At zero magnetic field, small
intensities, $T_c \approx \tau_{sc}$ and $N_1 \approx N_c$ one has
from equations~(\ref{sigmapsi}), (\ref{szsigma})
\[
\Sigma = \frac{(P_i G^{\rm opt}
T)^2}{n N_1} \frac{T_c}{\tau (1 - \eta)^2} \approx P_i^2 \frac{G^{\rm opt} T^2 T_c}{N_c \tau^{*2}} \approx
P_i^2 \frac{G^{\rm opt} \tau_{sc}}{N_c}
\]
and, thus, the critical value of $G^{\rm{opt}}$ is given by
\begin{equation} \label{critical}
G^{\rm{opt}}_{\rm cr,\,circ} = \frac{N_c}{P_i^2 \tau_{sc}} \:.
\end{equation}
It follows from experimental data presented in
figure~\ref{SDRratio}b that the spin filter effect develops to
full scale at $J = 75$\,mW, so that $G^{\rm {opt}}_{\rm cr,\,circ}
\sim$ $3 \times 10^{25}$\,$\textrm{cm}^{-3} \textrm{s}^{-1}$.
Taking $P_i = 0.24$, $\tau_{sc}$ = 700\,ps and $T = \tau^*$ (since
$\tau_s \gg \tau^*$) we obtain from equation~(\ref{critical}) the
estimation $N_c = 10^{15}$\,cm$^{-3}$. The best agreement with
experiment, see the details below, is obtained for the density of
deep paramagnetic centers $N_c = 3 \times 10^{15}$\,cm$^{-3}$, in
agreement with the crude estimation. It is worth to mention that,
according to equation~(\ref{pizero}), the critical generation rate
$G^{\rm{opt}}_{\rm cr,\,lin}$ at which the system exhibits
nonlinear behaviour under linearly polarized photoexcitation is
given by the condition $X=1$ with $X$ defined by
equation~(\ref{XYZ}) or, equivalently, by
\begin{equation} \label{criticallin}
G^{\rm{opt}}_{\rm cr,\,lin} = \frac{N_c}{\tau^*} \:.
\end{equation}

\section{Comparison with experiment and discussion} \label{sec:4}
One of the most convincing signatures of spin-dependent
recombination is dependence of the PL intensity on the light
polarization at zero magnetic field, or more precisely, an
increase in the PL intensity under the switching from linear to
circular polarization of the exciting light. In
figure~\ref{SDRratio} full squares show the measured SDR ratio
defined by $K_{\rm SDR} = I({\rm circ},B=0)/I({\rm lin},B=0)$,
where $I({\rm circ},B=0)$ and $I({\rm lin},B=0)$ are the PL
intensity under circularly- and linearly-polarized
photoexcitation. At very weak pumping power, the intensities
$I({\rm circ})$ and $I({\rm lin})$ coincide. With increasing the
power the PL intensity $I({\rm circ},B=0)$ is enhanced as compared
with $I({\rm lin},B=0)$ and, at $J > 100$\,mW, the SDR ratio
reaches values as high as 8 and even higher. A transverse magnetic
field of 5\,kG strength eliminates the enhancement caused by the
circular polarization of the incident light and, as a result, the
modified SDR ratio $K_{\rm SDR}^{(B= 5\,{\rm kG})} = I({\rm circ},
B=0)/I({\rm circ}, B=5\,{\rm kG})$ behaves as a function of $J$
identically to the dependence $K_{\rm SDR}(J)$. Simultaneously the
degree of PL circular polarization induced under
circularly-polarized photoexcitation (full circles in
figure~\ref{SDRratio}) monotonously increases up to 25\% which is
another important manifestation of the spin-dependent
recombination.
\begin{figure}
  \centering
    \includegraphics[width=0.8\linewidth]{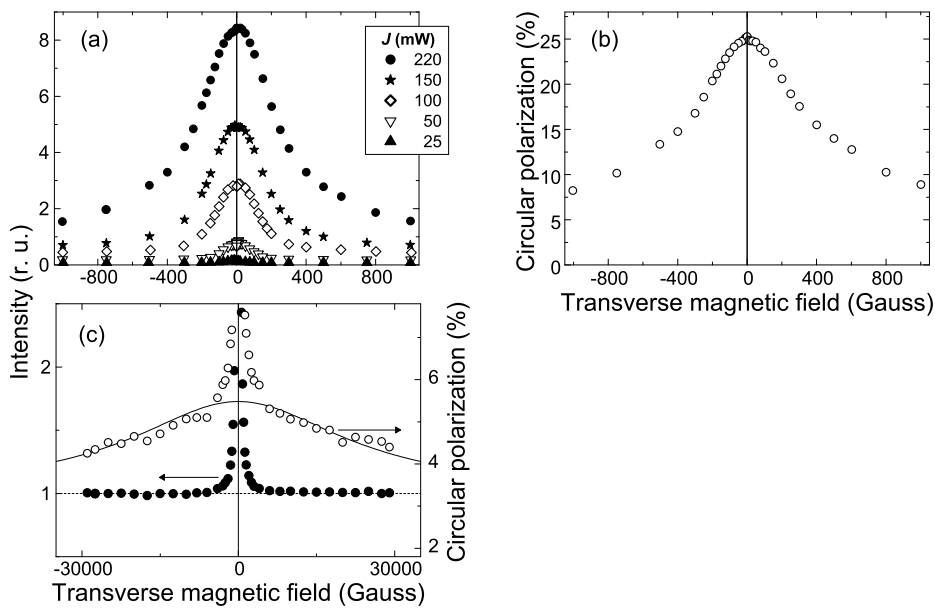}
    \caption{ The photoluminescence intensity (a,~c) and circular polarization (b,~c) as functions of a
    transverse magnetic field measured in GaAs$_{0.979}$N$_{0.021}$
for different powers of the circularly-polarized pumping.
Experimental points presented in panel (a) are obtained for the
following incident-light intensity $J$ (in mW): $\blacktriangle$
-- 25, $\bigtriangledown$ -- 50, $\diamond$ -- 100, $\star$ --
150, $\bullet$ -- 220. Panel (b) shows the upper part of the Hanle
curve at $J=220$\,mW and panel (c) shows the dependencies $I(B)$
(solid circles) and $\rho(B)$ (open circles) measured at
$J=220$\,mW in the wide range of magnetic fields. Solid line in
panel (c) is the Lorentzian $\rho(B)=\rho_{0}/(1+B^2/B_{1/2}^2) +
\rho_{res}$ with $B_{1/2}=25000$\,G. $\hbar\omega_{\rm
exc}=1.312$\,eV, $\hbar\omega_{\rm det}=1.17$\,eV.}
    \label{IntHanle2}
\end{figure}

Figure \,\ref{IntHanle2} shows typical curves of
magnetic-field-induced quenching of the photoluminescence (panels
(a) and (c)) and its depolarization (panels (b) and (c)) in
GaAs$_{0.979}$N$_{0.021}$ measured under circularly-polarized
pumping. One can see that the depolarization curve (Hanle effect)
is composed of narrow and wide contours with the half-widths
differing by two orders of magnitude. The main variation of narrow
Hanle contour takes place at the same magnetic field as for the
switching off the PL intensity enhancement which is a clear
evidence that the narrow and wide parts of the Hanle curve are
related to the spin depolarization of bound and free electrons,
respectively.

For a quantitative description of the data presented in
figures~\ref{SDRratio} and \ref{IntHanle2} we apply the
two-charge-state model for the selected set of parameter values,
i.e.,
\begin{eqnarray} \label{parameters1}
g = 1\:,\hspace{7 mm}g_c = 2\:,\hspace{7 mm}\tau^* = (c_n N_c)^{-1} = 1.9\,\textrm{ps},\\
\tau^*_h = (c_p N_c)^{-1} = 27 \,\textrm{ps}, \hspace{7 mm}\tau_s
= 140\,\textrm{ps}, \hspace{7 mm}\tau_{sc}=700\,\textrm{ps}\:.
\nonumber
\end{eqnarray}
Three remaining parameters of the theory, $P_i, P'$ and $N_c$, are
determined from comparison with the experimental data displayed in
figures~\ref{SDRratio} and \ref{IntHanle2}. According to
equations~(\ref{rhoexp}) and (\ref{Pprime}) values of $P_i$ and
$P'$ can be found as soon as the effective ``residual''
polarization $\rho_{\rm res}$ is estimated. Unfortunately, the
range of used magnetic field is insufficient to determine
$\rho_{\rm res}$ unambiguously: its value clearly lies in the
interval between 0 and 3.3\%. Taking into account that, as follows
from figure~\ref{SDRratio}(b), $\rho_{\rm exp}(B=0, {\rm high}~J)
\approx 25\%$ and $\rho_{\rm exp}(B=5\,\textrm{kG}) \approx 6.7\%$
and using equations~(\ref{rhoexp}), (\ref{Pprime}) we obtain $P'
\approx 0.25$ and $P_i \approx 0.27$ (if $\rho_{\rm res} = 0$) and
$P' \approx 0.22$ and $P_i\approx0.16$ (if $\rho_{\rm res} =
3.3\%$). An additional variation of these values follows from the
difference of $P(B=0, {\rm high}~ J)$ from unity since the ratio
$\tau^*/\tau_s$ is small but finite. The self-consistent
calculation, see figure~\ref{P(Pi)}, shows that one should
substitute into equation~(\ref{Pprime}) the value $P(B=0, {\rm
high}~ J) = 0.9$ so that the values $P' = 0.25, P_i = 0.27$ are
replaced by $P' = 0.28,  P_i = 0.24$. Due to the same reason, the
values of $P', P_i$ corresponding to $\rho_{\rm res} = 3.3\%$
should also be changed from $0.22, 0.16$ to $0.27, 0.14$. The
curves in figure~\ref{SDRratio} are calculated for the two pairs
of $P', P_i$ values with $N_c$ being the final fitting parameter.
One can see that the set
\begin{equation} \label{parameters2}
P_i = 0.24, \hspace{5 mm}P' = 0.28\hspace{5 mm} {\rm and}
\hspace{5 mm} N_c = 3 \times 10^{15}\,{\rm cm}^{-3}
\end{equation}
rather well simulate the experimental dependencies.

\begin{figure}
  \centering
    \includegraphics[width=0.45\linewidth]{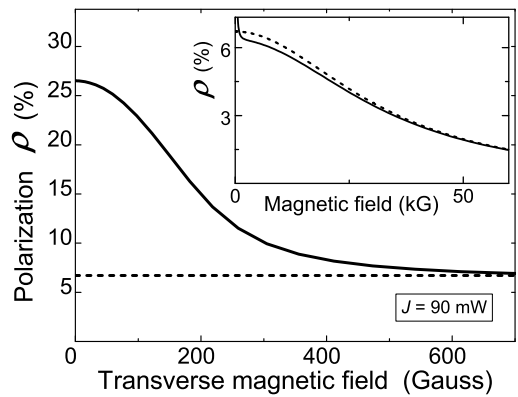}
  \caption{Calculated Hanle curve for the free-electron spin polarization $\rho$ in a coupled
  spin system of free and bound photoelectrons. Inset shows the larger magnetic-field range.
  Solid curve represents the exact calculation while dashed curve is
obtained assuming unpolarized bound electrons in which case the spin filter is switched off.}\label{f_valleys}
\end{figure}
The complete set (\ref{parameters1}), (\ref{parameters2}) of
parameters allows us to simulate the effect of transverse magnetic
field and analyze consequences of variation of each parameter. The
theoretical Hanle curve calculated at pump power $J=90$\,mW is
shown in figure~\ref{f_valleys} in the narrow range up to 700\,G
and, in inset, in the wide range up to 60\,kG. As mentioned in
section~\ref{sec:2}, the Hanle curve outside the narrow contour is
described by the Lorentzian (\ref{arbint}). This property is
demonstrated by drawing a subsidiary (dashed) line calculated
assuming formally that paramagnetic centers are kept unpolarized,
i.e., setting ${\bm S}_c = 0$ in
equations~(\ref{7IIa})--(\ref{7IIe}). Then the densities $n, N_1$
and $p$ are independent of the magnetic field and the polarization
$\rho(B)$ is indeed given by the Lorentzian (\ref{arbint}). In
figure~\ref{f_valleys} the exact (solid) curve is rather well
approximated by the dashed curve for $B > 0.5$\,kG.

\begin{figure}
  \centering
    \includegraphics[width=0.4\linewidth]{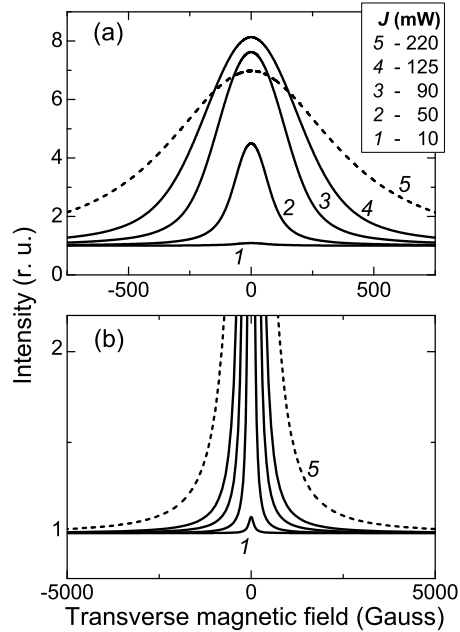}
  \caption{ Magnetic-field dependence of the PL intensity $I(B)$ calculated at
  different powers $J$ of circularly polarized excitation and presented on small (a) and large (b) field
  scales. The curves 1$\dots$5 correspond to the following values
  of $J$: $1 - 10$\,mW, $2 - 50$\,mW, $3 - 90$\,mW,  $4 - 170$\,mW, $5 - 220$\,mW.
  } \label{Int(B)}
\end{figure}
The two-charge-state model rather well reproduces manifestation of
the Hanle effect in the PL intensity. The dependence of PL
intensity on the transverse magnetic field calculated for
different $\sigma$-pump intensities is shown in
figure~\ref{Int(B)}. In agreement with experiment the intensity
$I$ decreases in relatively weak fields $|B| < 5$\,kG which
suppress the polarization of bound electrons. Moreover, half-width
of the decaying part of the function $I(B)$ is close to that of
the narrow contour of the curve $\rho(B)$. At high fields $|B| >
5$\,kG gradually depolarizing the free electrons, the PL intensity
is field-independent, see figure~\ref{Int(B)}(b), in accordance
with experimental data of figure~\ref{IntHanle2}(c).

Figure~\ref{P(Pi)} displays the relation between the average
electron spin polarization $P$ and the initial polarization $P_i$
for different pumping powers $J$. At low values of $J$ this
relation is described by equation~(\ref{lowintb}) which can be
rewritten in the form
\begin{equation} \label{PtoPi}
P = \left( c_0 + c_1\, \frac{J}{1\,{\rm mW}}\right) P_i \:,
\end{equation}
where $c_0 = T/\tau^*$ and $c_1 = c_0^2 \tau_{sc} G^{\rm
opt}(1\,{\rm mW})/N_c$. Thus, at weak pumping and small $P_i$, the
spin polarization $P$ is proportional to $P_i$ with the
proportionality coefficient being a linear function of $J$. This
approximate relation is shown in figure~\ref{P(Pi)} by dotted
straight lines. One can see that, at small $J$, the onset slopes
of the exact and approximate curves $P(P_i)$ coincide; at moderate
pumping the slopes slightly deviate, and only at large values of
$J$ the difference between slopes gets remarkable. With increasing
$P_i$ the dependence $P(P_i)$ at finite fixed $J$ becomes
sublinear and finally saturates to $\tau_s/(\tau + \tau_s) \approx
1$.
\begin{figure}
  \centering
    \includegraphics[width=0.5\linewidth]{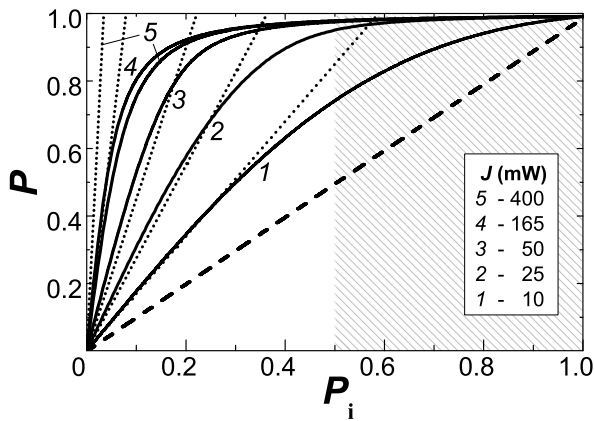}
    \caption{Relation between average spin polarization, $P$, of the conduction photoelectrons
    and their initial spin polarization $P_i$ calculated for different excitation powers varying
    from 10\,mW to 400\,mW. The dashed line shows
    the limiting relation $P = P_i \tau_s/(\tau + \tau_s) \approx P_i$ attainable
    at vanishingly small intensity where the spin filter is switched off. The dotted straight
    lines represent the linear approximation (\ref{PtoPi}). The shaded area $P_i > 0.5$ is unachievable
    in a bulk undeformed semiconductor. }
      \label{P(Pi)}
      \end{figure}

\begin{figure}
  \centering
    \includegraphics[width=0.8\linewidth]{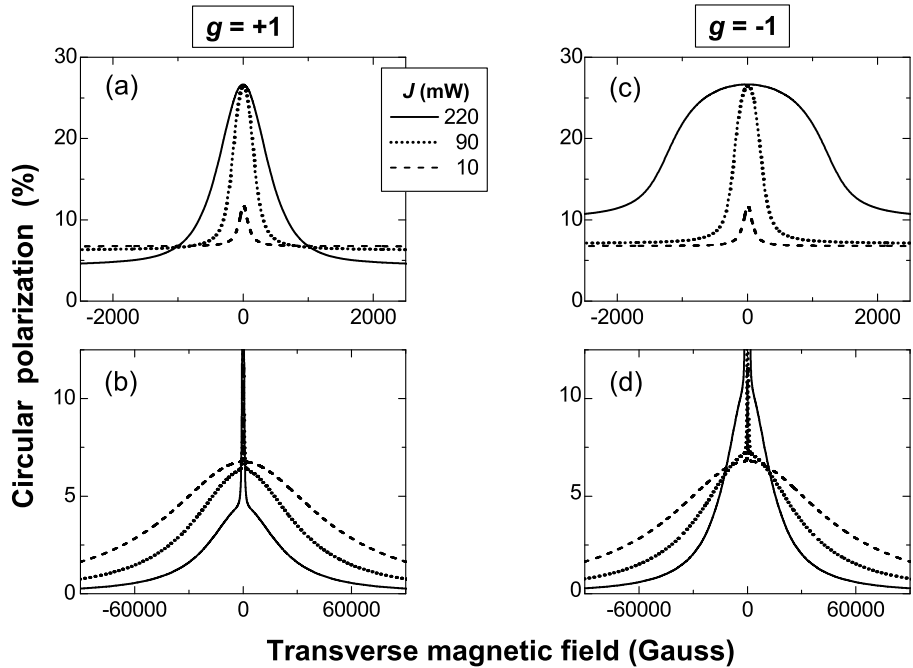}
    \caption{Hanle curves calculated for positive, $g=+1$ (a, b), and negative, $g=-1$ (c, d),
    values of the conduction electron $g$ factor and for different excitation powers $J=10$\,mW (dashed),
    90\,mW (dotted) and 250\,mW
    (solid). Other parameters are indicated in text, see (\ref{parameters1}) and (\ref{parameters2}).    }
    \label{gSign}
\end{figure}

Figure \ref{gSign} illustrates effect of the sign of free-electron
$g$ factor on the Hanle curve. The curves in the left and right
panels are calculated for $g=1$ and $g=-1$, respectively, while
the parameter $g_c=2$ is kept constant. One can see that, the
value of $\rho(B=5\,\textrm{kG})$ (which determines the amplitude
of the wide Hanle contour) calculated at $g=1$ remarkably
decreases with the increasing $J$. In contrast, at $g=-1$ the wide
contour moves upwards as the intensity $J$ increases from 10\,mW
to 250\,mW and, moreover, the narrow contour becomes much wider as
compared to the case of $g=1$. This can be understood taking into
account that, according to equation~(\ref{sxscx}), in the system
of coupled free and bound electron spins they precess in the
transverse magnetic field faster if the signs of $g$ and $g_c$
coincide and slower if the signs of $g$ and $g_c$ are opposite. As
a result, for negative $g$ and positive $g_c$ the function $\Psi$
in equation~(\ref{psi}) becomes more robust to the influence of
the transverse magnetic fields in the region $B\le$ 2\,kG.
Comparison of the experimental Hanle curves in figure~2 of
reference~\cite{physica} with figures~\ref{gSign}(b) and (d)
confirms the positive sign of $g$ determined in
GaAs$_{0.979}$N$_{0.021}$ under oblique pumping and
detection~\cite{SST2008}.

\begin{figure}
  \centering
    \includegraphics[width=0.8\linewidth]{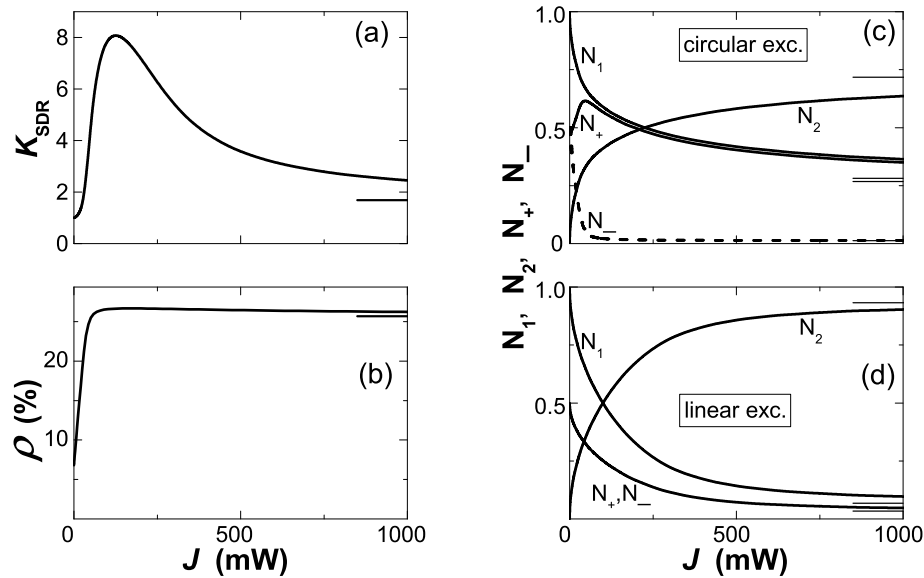}
    \caption{Variation of $K_{\rm SDR}$, $\rho$, $N_1 / N_c$, $N_2 / N_c$, $N_+ / N_c$
    and $N_- / N_c$ with the increasing pump intensity $J$ calculated in the wide range of $J$ for the set of
    parameters (\ref{parameters1}) and (\ref{parameters2}). Short horizontal lines indicate
    the limiting values at $J \to
    \infty$.}
      \label{KRoN1N2(G)}
      \end{figure}

The behaviour of the system under study in the wide range of pump
intensities is demonstrated in figure~\ref{KRoN1N2(G)}. At weak
intensities the dependence of $N_1$ and $N_2$ is insensitive to
the polarization of incident light: $Y= N_2/N_c = (N_c - N_1)/N_c
\propto \sqrt{J}$. With increasing $J$ the dependence $N_2(J)$
tends to saturation. For $\tau^* = 1.9$\,ps, $\tau_{sc}$ = 700\,ps
and $P_i = 0.24$, we have for the critical values
(\ref{critical}), (\ref{criticallin}) of the generation rates:
$G^{\rm{opt}}_{\rm cr,\,lin} \gg G^{\rm{opt}}_{\rm cr,\,circ}$.
Therefore, the deviation from the square-root law begins earlier
for the circular polarization. The limiting values of $Y$ at
$G^{\rm{opt}} \to \infty$ are the solution $Y_{\infty}({\rm circ})
= 0.72$ of equations~(\ref{yinfty}) and the solution
$Y_{\infty}({\rm lin}) = c_n/(c_n + c_p) = 0.93$ of
equation~(\ref{pizero}) at $X \to \infty$. The limiting values of
$K_{\rm SDR}$ and $\rho$ are $[(1 + a)Y_{\infty}({\rm circ})]^{-2}
= 1.7$ and $(P' P_i \tau_{sc}/\tau^*_h) Y_{\infty}({\rm circ}) =
0.26$, respectively. The free-electron spin polarization tends to
$P_{\infty} = (P_i \tau_{sc}/\tau^*_h) Y_{\infty}({\rm circ}) =
0.9$. Note, however, that these limiting values are achieved at
very high intensities far exceeding $J=1000$\,mW where the model
defined by the parameter set (\ref{parameters1}),
(\ref{parameters2}) is scarcely applicable. Particularly, an
increase in the photoelectron and photohole densities at very high
generation rates can lead to a remarkable contribution of the
interband electron-hole recombination and the term $\gamma
n_{\pm}p$ in equation~(\ref{maineqs}) should be taken into
consideration. A nonmonotonic behavior of $K_{\rm SDR}$ as a
function of $J$ is an interesting and intuitively unexpected. In
figure~\ref{SDRratio} experimental points only begin to show such
behaviour. However, references~\cite{Int1,Int2} report observation
of a maximum in the dependence $K_{\rm SDR}(J)$. As one can see
from figure~\ref{KRoN1N2(G)}(b), the degree of PL circular
polarization rapidly increases with increasing $J$ and reaches a
maximum value $\rho_{\rm max} = 0.27$, then it almost stops to
depend on $J$ slowly decreasing to $\rho_{\infty}$ = 0.26. The
corresponding values of $P$ are $P_{\rm max}$ = 0.94 and
$P_{\infty}$= 0.90.





\section{Conclusion} \label{sec:5}
In summary, we have developed a spin-dependent theory of
Shockley--Read recombination for three- and two-charge-state
defects. The two-charge-state model characterized by nine
parameters satisfactorily describes the coupled spin system of
free and bound photoelectrons in undoped GaAsN alloys. In the
weak- and strong-pumping limits, we present simple analytic
equations which allow us to predict many properties of the coupled
system not applying to computer simulations. Here we have
concentrated on cw optical excitation. The further step could be a
comprehensive analysis of time-resolved kinetics, including PL
decay after short-pulse photoexcitation and pump-probe
spectroscopy. It might be also interesting to study the
strong-pumping regime where the recombination
of free electrons and holes
acquires an important role in the kinetics of photocarriers and
work out predictions of specific behaviour of the system in this
regime. Experimentally, it would be rather instructive to
demonstrate unambiguously an existence or nonexistence of the
deep-center level in GaAsN with no electron on the center.

\ack{This research was supported by the programmes of the Russian
Academy of Sciences, Russian Foundation for Basic Research and
TARA project of the University of Tsukuba. We are grateful to
A~Yu~Egorov and V~M~Ustinov for useful discussions.}

\end{document}